\documentstyle[12pt,epsf]{article}
\newcommand{\beq}{\begin{eqnarray}}
\newcommand{\eeq}{\end{eqnarray}}
\begin{document}
\title{Bose-Einstein condensation of atomic gases in a harmonic oscillator confining potential trap}
\author{Klaus Kirsten    \cite{kk}\\University of Leipzig\\
Institute of Theoretical Physics\\
Augustusplatz 10, 
04109 Leipzig\\
\\and\\
\\
David J. Toms   \cite{djt}          \\
Department of Physics\\ University of Newcastle Upon Tyne,\\
Newcastle Upon Tyne, U. K. NE1 7RU}

\date{\today}
\maketitle

\begin{abstract}
We present a model which predicts the temperature of Bose\--Ein\-stein condensation in atomic alkali gases and find excellent agreement with recent experimental observations. A system of bosons confined by a harmonic oscillator potential is not characterized by a critical temperature in the same way as an identical  system which is not confined. We discuss the problem of Bose-Einstein condensation in an isotropic harmonic oscillator potential analytically and numerically for a range of parameters of relevance to the study of low temperature gases of alkali metals. 
\end{abstract}
\eject

A stunning development in the last year has been the demonstration of Bose-Einstein condensation (BEC) in gases of rubidium \cite{rub}, lithium \cite{lith}, and most recently sodium \cite{sod}. This experimental work has stimulated theoretical studies to try to understand the underlying physics of the situation \cite{BaymPethick,Fetter,DolString}. For many purposes the complicated magnetic traps used in the experiments can be approximated by harmonic oscillator potentials. There have been several studies of BEC in harmonic oscillator confining potentials \cite{oldone,MITcrowd,PhysLett,HaugRav}. The purpose of our paper is to examine the condensation of bosons in a harmonic oscillator potential in a detailed way, and to point out a serious flaw in the analyses of Refs. \cite{MITcrowd,PhysLett,HaugRav}; namely, unlike the situation for a boson gas with no external confining potential \cite{HuangPathria}, there does not exist a critical temperature which signals a phase transition. However, we will show that there is a temperature at which the specific heat has a maximum which can be identified as the temperature at which BEC occurs. By using parameters relevant to the experiments with alkali gases \cite{rub,lith,sod}we find excellent agreement with the experiments.

The fact that a gas of bosons (neglecting interactions) in a harmonic oscillator potential does not have a phase transition at some critical temperature is already apparent from the early work of \cite{oldone}. (This will also be shown below in a very simple way.) A similar situation occurs for a system of charged bosons in a homogeneous magnetic field; in three spatial dimensions BEC does not occur in the same way as for the case where there is no magnetic field \cite{Schafroth}. The same is true for bosons confined by spatial boundaries \cite{Pathria}. A general criterion to decide whether or not BEC occurs has been given recently by us \cite{KKDJT}, and it is easy to show that the criterion is not met for a system of bosons confined by a harmonic oscillator potential.

Since the experiments of Refs. \cite{rub,lith,sod} obviously are seeing BEC, a natural question which arises concerns the exact nature of the phenomenon. If BEC as found normally for the free boson gas is impossible for a system of bosons in a confining potential, then in what sense does BEC occur ? A natural criterion which has been used in systems of finite size has been to look at the maximum of the specific heat \cite{PathPak}. We will apply this criterion to the case of bosons confined by a harmonic oscillator potential.

The simple model which we will discuss here is appropriate to neutral spin-0 bosons confined by a harmonic oscillator potential. For mathematical simplicity we will only deal with the case of an isotropic potential in this paper. ( The more general analysis of anisotropic potentials, along with mathematical details relevant to the analysis presented here will be given elsewhere \cite{KKDJTinprep}.) We will assume that the system can be described by a grand canonical ensemble. The grand potential is defined by
\beq
q=-\sum_N\ln\left(1-z\exp(-\beta E_N)\right)\label{eq1}
\eeq
where $\beta=(kT)^{-1}$, $E_N$ are the energy levels, and $z=e^{\beta\mu}$ is the fugacity in terms of the chemical potential $\mu$. It proves convenient to expand the logarithm in (\ref{eq1}) to obtain
\beq
q=\sum_{n=1}^{\infty}\frac{z^n}{n}\sum_N\exp(-n\beta E_N)\;.\label{eq2}
\eeq
For an isotropic harmonic oscillator characterized by an angular frequency $\omega$ the energy levels are given by $E_k=(k+3/2)\hbar\omega$ with multiplicity $(k+1)(k+2)/2$ where $k=0,1,2,\ldots$. The sum over $N$ in (\ref{eq2}) may be performed to obtain
\beq
q=\sum_{n=1}^{\infty}\frac{e^{n\beta(\mu-3/2\hbar\omega)}}{n(1-e^{-nx})^{3}}\label{eq3}
\eeq
where we have defined the dimensionless variable $x=\hbar\omega/(kT)$. The number of particles is given by $\displaystyle{N=\beta^{-1}\left(\frac{\partial q}{\partial\mu}\right)_{T,\omega}}$, which becomes
\beq
N=\sum_{n=1}^{\infty}\frac{e^{n\beta(\mu-3/2\hbar\omega)}}{(1-e^{-nx})^3}\;.\label{eq4}
\eeq

In order that the number of particles remain positive, it is necessary for $\mu\le3/2\hbar\omega$. (More generally, we require $\mu\le E_0$ where $E_0$ is the lowest energy level.) Normally the critical temperature for BEC is the temperature at which $\mu=E_0$, which for the isotropic harmonic oscillator reads $\mu=3/2\hbar\omega$. It is now easy to see that BEC cannot occur in the same way for bosons confined in the harmonic oscillator potential as it does for bosons in free space. In the case of the free boson gas in free space with no confining potential, as the temperature is lowered the chemical potential $\mu$ increases from negative values towards the value 0. (This is in agreement with the general result $\mu=E_0$ quoted above since the lowest energy level is zero for the free boson gas.) The value of the temperature at which $\mu=0$ defines a critical temperature $T_c$ determined in terms of the particle density. At temperatures lower than $T_c$, $\mu$ remains frozen at the value $\mu=0$, and the number of particles found in excited states is bounded. If the total number of particles exceeds this bound then the only possibility is for the excess particles to be found in the ground state, giving rise to BEC. This standard scenario is described in \cite{HuangPathria} in some detail. 

In the case of (\ref{eq4}) it is easy to see that the number of particles satisfies the following simple bound (assuming $\omega\ne0$)
\beq
N>\frac{e^{\beta(\mu-3/2\hbar\omega)}}{1-e^{\beta(\mu-3/2\hbar\omega)}}\;.\label{eq5}
\eeq
As $\mu\rightarrow3/2\hbar\omega$, which should signify the onset of BEC, this expression shows that $N$ increases without bound, unlike the free boson gas where as $\mu\rightarrow0$ $N$ has an upper bound. To phrase this another way, no matter how low $T$ becomes (so long as it remains non-zero) it is always possible to solve (\ref{eq4}) for $\mu$ with $\mu<3/2\hbar\omega$ for any finite value of $N$ no matter how large $N$ is. If $N$, as given in (\ref{eq4}) had an upper bound, as for the case of the free boson gas in three spatial dimensions, this would no longer be the case. There is therefore a qualitative as well as a quantitative difference between the free boson gas and a system of bosons confined in a harmonic oscillator potential.

We now come to our analysis of (\ref{eq3}) and (\ref{eq4}) and our fundamental difference with earlier treatments \cite{MITcrowd,PhysLett,HaugRav}. We have treated, and will continue to treat, the energy spectrum for bosons confined in the harmonic oscillator potential as discrete. Refs.~\cite{MITcrowd,PhysLett,HaugRav} make the physically plausible but mathematically unsound assumption that if $x=\hbar\omega/(kT)<<1$, then it is possible to replace the sums which arise by integrals. It is precisely this procedure which leads to the erroneous conclusion that a critical temperature exists at which (in our units) $\mu=3/2\hbar\omega$. Obviously if the original sum (\ref{eq4}) has no upper bound as $\mu\rightarrow3/2\hbar\omega$, any approximation which leads to an upper bound is suspect. If the behaviour for small $x$ is desired, a safer approach is to take the exact results expressed in terms of sums, and to try to obtain expansions valid for small $x$. This is also useful for numerical purposes since the original sums (\ref{eq2}) and (\ref{eq4}) are not very rapidly convergent if $x$ is small.
For sodium \cite{sod}, if we take $T\simeq2\times10^{-6}$K, and take $\omega$ to be given by the geometric mean of the frequencies of 235 Hz, 410 Hz, and 745 Hz (i.e. $\omega/2\pi$=416 Hz), then $x$ is of order $10^{-2}$. Thus the behaviour of the thermodynamic quantities for small $x$ is of interest. 

In addition to the dimensionless variable $x$ defined earlier, it is convenient to define a new dimensionless variable $\epsilon$ related to the chemical potential $\mu$ by $\mu=\hbar\omega(3/2-\epsilon)$. If we assume that $x<<1$ and $\epsilon<<1$, then the leading behaviour for $q$ in addition to $N$ is well accounted for by approximating $e^{-n\epsilon x}(1-e^{-nx})^{-3}\simeq (nx)^{-3}$ which is obtained by a simple Taylor expansion. (For the anisotropic oscillator this also shows that identifying $\omega$ with the geometric mean of the frequencies is justified for the leading order term.) This approximation can be criticized on the grounds that no matter how small we take $x$, $n$ eventually becomes large enough so that $nx>>1$. A more satisfactory procedure from the mathematical standpoint is to convert the sums defining $q$ and $N$ into contour integrals, and to deform the contour in an appropriate way for obtaining at least an asymptotic expansion of the desired function. The details of this procedure are somewhat involved and will be given elsewhere \cite{KKDJTinprep}. 

If $x<<1$ and $\epsilon<<1$, then it can be shown that
\beq
N\simeq\zeta(3)x^{-3}+(\epsilon x)^{-1}+\frac{3}{2}\zeta(2)x^{-2}-\frac{\ln x}{x}-\epsilon\zeta(2)x^{-2}\label{eq6}
\eeq
is a good approximation. ($\zeta(z)$ denotes the Riemann $\zeta$-function.) It is possible to obtain an expansion for $q$ to the same level of accuracy; however, the expression involves more complicated functions than that for $N$ and will not be given here. It can be noted that the leading term in (\ref{eq6}) agrees with that found using the naive procedure mentioned in the previous paragraph. It is possible to obtain an expression for the specific heat $C$ valid for $x<<1$ and $\epsilon<<1$~:
\beq
k^{-1}C\simeq12\zeta(4)x^{-3}+\frac{27}{2}\zeta(3)x^{-2}\;.
\label{eq7}
\eeq
The internal energy is given by
\beq
U\simeq\hbar\omega\left\lbrace3\zeta(4)x^{-4}+\frac{9}{2}
\zeta(3)x^{-3}+\frac{13}{4}\zeta(2)x^{-2}+\frac{3}{2}
(\epsilon x)^{-1}-3\epsilon\zeta(3)x^{-3}\right\rbrace\;.
\eeq
These expressions agree to a good approximation with numerical results found by analyzing the exact sums when $x<<1$ and $\epsilon<<1$ hold. In particular, the result for $N$ in (\ref{eq6}) is useful for determining the chemical potential in terms of the number of particles and the temperature.

For large $x$ (in fact if $x\ge1$) the original sums (\ref{eq3}) and (\ref{eq4}) converge quite rapidly and are useful for numerical computation. For $x>>1$ the specific heat becomes 
\beq
k^{-1}C\simeq 3x^2e^{-x}\;.
\eeq
The specific heat therefore vanishes as $x\rightarrow\infty$ corresponding to $T\rightarrow0$. It is also possible to obtain analytic approximations valid for $x<<1$ but without assuming $\epsilon<<1$. These expressions all involve more complicated functions than those we have presented here and will not be given.

Unfortunately it is difficult to find a reliable analytic approximation for the specific heat which allows us to study whether it has a maximum, and if it does, at what temperature it occurs. Instead we have studied the specific heat numerically by doing the relevant sums for a range of temperatures. It is necessary to solve (\ref{eq4}) for the chemical potential in terms of $N$. When this is done for the case $N=5\times10^5$ \cite{sod}, it is found that over a very small range of $x$, corresponding to a very small range of temperatures, $\epsilon$ changes from large values of the order of 100 to small values of the order $10^{-3}$. This is a good indicator that the specific heat has an unusual behaviour in this temperature range. We have shown our results for the specific heat over the range of interest in Fig.~1.
\begin{figure}
\begin{center}
\leavevmode
\epsffile{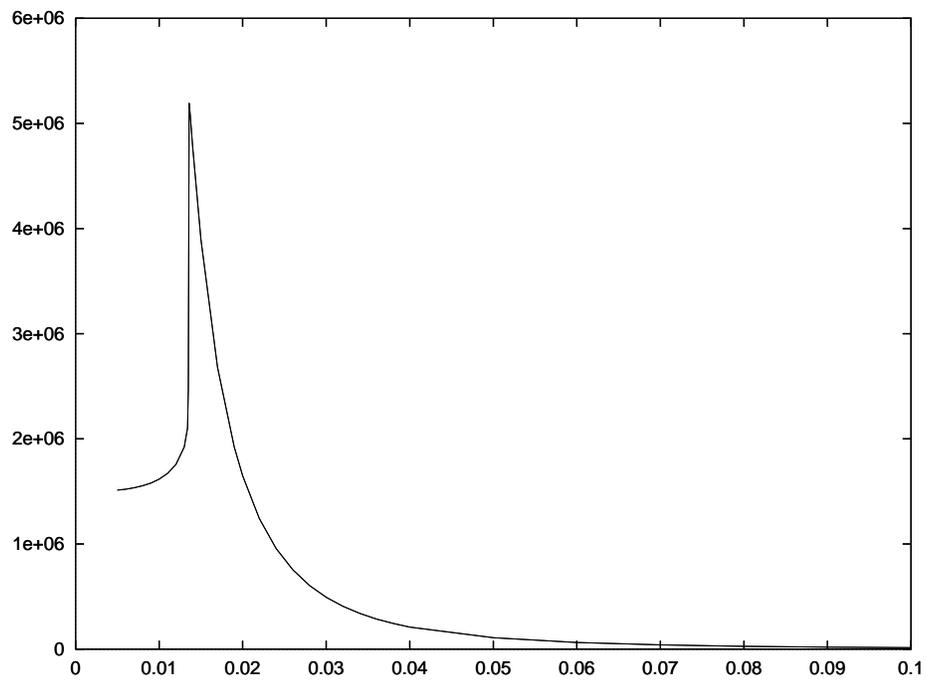}
\end{center}
\caption{The specific heat in units of $k$ as a function of $x=\hbar\omega/(kT)$.}
\end{figure}

The ordinate gives the specific heat in units of the Boltzmann constant $k$, and the abscissa is $x=\hbar\omega/(kT)$ with $\omega/2\pi=416$ Hz. It is readily apparent from this figure that the specific heat undergoes a very rapid change over a very small range of temperatures. The maximum in the specific heat occurs for $x\simeq0.0136$ which corresponds to a temperature of $T\simeq1.47\times10^{-6}$ K. This is in remarkable agreement with the value of $2\times10^{-6}$ K in Ref.~\cite{sod}. It supports our proposal that the BEC observed in sodium gas at least is associated with the maximum in the specific heat rather than with a phase transition as occurs for the free boson gas.

The specific heat is perfectly continuous and smooth at its maximum. We have shown the behaviour of the specific heat over a small range of $x$ where it changes most rapidly in Fig.~2. Our results contrast markedly with the discontinuous behaviour found in Refs.~\cite{MITcrowd,PhysLett,HaugRav} which was found by approximating sums with integrals. This further supports our claim that such an approximation leads to unreliable conclusions.

\begin{figure}
\begin{center}
\leavevmode
\epsffile{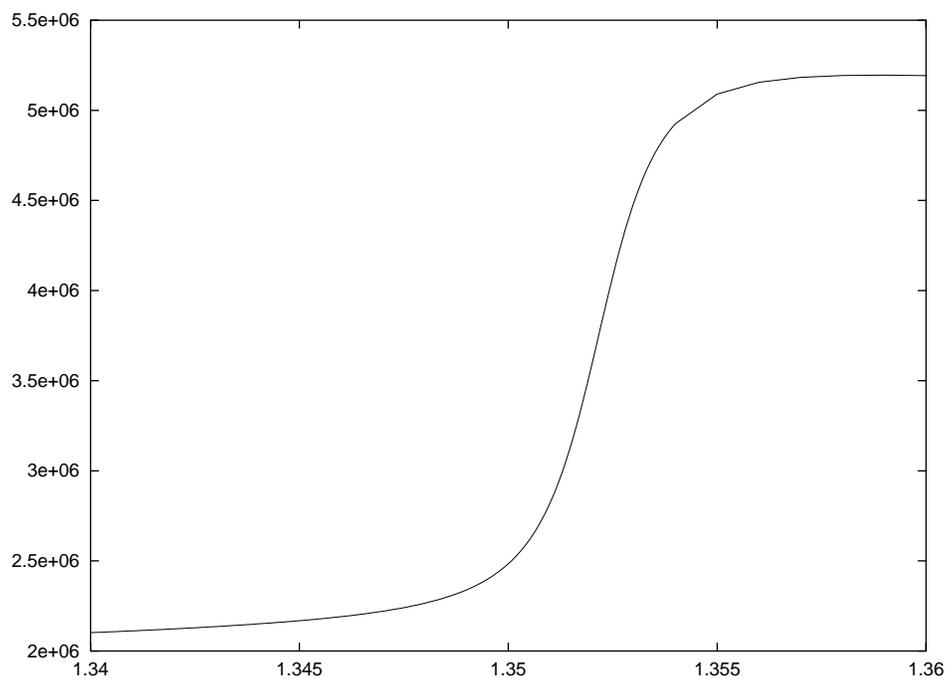}
\end{center}
\caption{The specific heat in units of $k$ as a function of $100x$ where $x=\hbar\omega/(kT)$.}
\end{figure}

Although we have concentrated on sodium, we can easily extend the analysis to rubidium \cite{rub} and lithium \cite{lith}. We have worked out the maximum in the specific heat for $N=2\times10^5$ and $N=2\times10^4$, and find that the maximum occurs at $x\simeq0.0185$ and $x\simeq0.0408$ respectively. (The precise graphs in these two cases are qualitatively the same as those in Figs.~1--2.) Using the oscillator frequencies given in Ref.~\cite{lith} for lithium, and taking $N=2\times10^5$ we find the temperature at which the specific heat has a maximum is $T\simeq380$ nK in good agreement with the range of 100--400 nK for the experiment. For rubidium with $N=2\times10^4$ we find $T\simeq71$ nK if we use the oscillator frequencies of Ref.~\cite{rub}. If we use the frequencies given in Ref.~\cite{BaymPethick} for the strong trap we find $T\simeq124$ nK, again in close agreement with experiment. Given the simplicity of our model, the predicted condensation temperatures are in remarkable agreement with experiment.

In conclusion, we have studied the condensation of spin-0 bosons confined by an isotropic harmonic oscillator potential. We have shown that although there is no phase transition such as that occurring in the free unconfined boson gas, it is possible to identify a temperature at which BEC occurs by looking at the maximum in the specific heat. Using parameters relevant to the experiments which have been performed on alkali gases, we find that our predicted temperatures are in very close agreement with experiment. Although the gases in the experiments are dilute, it would be of interest to include weak interactions among the atoms, since interactions have proven important in other contexts \cite{BaymPethick,Fetter,DolString}. Our results seem to indicate that the transition temperature is not affected much by interactions. In any case, it would not be prudent to use the free boson gas as the zeroth order approximation in any perturbative treatment of interactions. Finally, it is of extreme interest to compare our theoretical predictions for the behaviour of the specific heat with experiment in a much more substantial way.

\vspace{0.75cm}\noindent{\bf Acknowledgements}\\

We would like to thank L.~D.~L.~Brown for helpful discussions about the numerical results.


\begin{thebibliography}{99}
\bibitem[\dag]{kk}
{\tt kirsten@tph100.physik.uni-leipzig.de}.
\bibitem[\ddag]{djt}
{\tt d.j.toms@newcastle.ac.uk}.
\bibitem{rub}
M.~H.~Anderson, J.~R.~Ensher, M.~R.~Matthews, C.~E.~Wieman, and E.~A.~Cornell, Science {\bf 269}, 198 (1995).
\bibitem{lith}
C.~C.~Bradley, C.~A.~Sackett, J.~J.~Tollett, and R.~G.~Hulet, Phys. Rev. Lett. {\bf 75}, 1687 (1995).
\bibitem{sod}
K.~B.~Davis, M.-O.~Mewes, M.~R.~Andrews, N.~J.~van Druten, D.~S.~Durfee, D.~M.~Kurn, and W.~Ketterle, Phys. Rev. Lett. {\bf 75}, 3969 (1995).
\bibitem{BaymPethick}
G.~Baym and C.~J.~Pethick, Phys. Rev. Lett. {\bf 76}, 6 (1996).
\bibitem{Fetter}
A. Fetter (unpublished).
\bibitem{DolString}
F.~Dalfovo and S. Stringari (unpublished).
\bibitem{oldone}
S.~R.~de Groot, G.~J.~Hooyman, and C.~A.~ten Seldam, Proc. Roy. Soc. (London) {\bf A 203}, 266 (1950).
\bibitem{MITcrowd}
V.~Bagnato, D.~E.~Pritchard, and D.~Kleppner, Phys. Rev. A {\bf 35}, 4354 (1987).
\bibitem{PhysLett}
S.~Grossmann and M.~Holthaus, Phys. Lett. A {\bf 208}, 188 (1995).
\bibitem{HaugRav}
H.~Haugerud and F.~Ravndal (unpublished).
\bibitem{HuangPathria}
See for example F. London, {\em Superfluids II\/}, (John Wiley, New York, 1954), K. Huang, {\em Statistical Mechanics\/} (John Wiley, New York, 1987) or R.~K.~Pathria, {\em Statistical Mechanics\/} (Pergammon, Oxford, 1972).
\bibitem{Schafroth}
M.~R.~Schafroth, Phys. Rev. {\bf 100}, 463 (1955).
\bibitem{Pathria}
See R.~K.~Pathria, Can. J. Phys. {\bf 61}, 228 (1983) for a review.
\bibitem{KKDJT}
K. Kirsten and D.~J.~Toms, Phys. Lett. B {\bf 368}, 119 (1996).
\bibitem{PathPak}
See H.~R.~Pajkowski and R.~K.~Pathria, J. Phys. A {\bf 10}, 561 (1977), as well as Ref. \cite{Pathria} for a discussion of this subject.
\bibitem{KKDJTinprep}
K.~Kirsten and D.~J.~Toms, paper in preparation.
\end{thebibliography}
\end{document}